\documentclass[aps,twocolumn]{revtex4}
\usepackage[utf8]{inputenc}
 \usepackage{amsmath,amssymb}
\usepackage{color}
\begin{document}
\title{Barrow Entropy Corrections to Friedmann Equations}
\author{Ahmad Sheykhi\footnote{asheykhi@shirazu.ac.ir}}
\address{Department of Physics, College of
Sciences, Shiraz University, Shiraz 71454, Iran\\
Biruni Observatory, College of Sciences, Shiraz University, Shiraz
71454, Iran}

 \begin{abstract}
Inspired by the Covid-$19$ virus structure, Barrow argued that
quantum-gravitational effects may introduce intricate, fractal
features on the black hole horizon [Phys. Lett. B {\bf808} (2020)
135643]. In this viewpoint, black hole entropy no longer obeys the
area law and instead it can be given by $S\sim A^{1+\delta/2}$,
where the  exponent $\delta$ ranges $0\leq\delta\leq1$, and
indicates the amount of the quantum-gravitational deformation
effects. Based on this, and using the deep connection between
gravity and thermodynamics, we disclose the effects of the Barrow
entropy on the cosmological equations. For this purpose, we start
from the first law of thermodynamics, $dE=TdS+WdV$, on the
apparent horizon of the Friedmann-Robertson-Walker (FRW) Universe,
and derive the corresponding modified Friedmann equations by
assuming the entropy associated with the apparent horizon has the
form of Barrow entropy. We also examine the validity of the
generalized second law of thermodynamics for the Universe enclosed
by the apparent horizon. Finally, we employ the emergence scenario
of gravity and extract the modified Friedmann equation in the
presence of Barrow entropy which coincide with one obtained from
the first law of thermodynamics. When $\delta=0$, the results of
standard cosmology are deduced.

\end{abstract}
 \maketitle

 \newpage
\section{Introduction\label{Intro}}

The fast spreading of Covid-$19$ virus around the world in $2020$
and its continuation until now in $2021$ provide strong
motivations for many scientists to consider the structure of this
virus from different perspectives. Inspired by the fractal
illustrations of this virus, recently Barrow proposed a new
structure for the horizon geometry of black holes \cite{Barrow}.
Assuming a infinite diminishing hierarchy of touching spheres
around the event horizon, he suggested that black hole horizon
might have intricate geometry down to arbitrary small scales. This
fractal structure for the horizon geometry, leads to finite volume
and infinite (or finite) area. Based on this modification to the
area of the horizon, the entropy of the black holes no longer
obeys the area law and will be increased due to the possible
quantum-gravitational effects of spacetime foam. The modified
entropy of the black hole is given by \cite{Barrow}
\begin{eqnarray}\label{S}
S_{h}= \left(\frac{A}{A_{0}}\right)^{1+\delta/2},
\end{eqnarray}
where $A$ is the black hole horizon area and $A_0$ is the Planck
area. The exponent $\delta$ ranges as $0\leq\delta\leq1$ and
stands for the amount of the quantum-gravitational deformation
effects. When $\delta=0$, the area law is restored and
$A_{0}\rightarrow 4G$, while $\delta=1$ represents the most
intricate and fractal structure of the horizon. Although the
corrected entropy expression (\ref{S}) resembles Tsallis entropy
in non-extensive statistical thermodynamics \cite{Tsa,Matin,Maj},
however, the origin and motivation of the correction, as well as
the physical principles are completely different. Some efforts
have been done to disclose the influences of Barrow entropy in the
cosmological setups. A holographic dark energy model based on
entropy (\ref{S}) was formulated in \cite{Emm1}. It was argued
that this scenario can describe the history of the Universe, with
the sequence of matter and dark energy eras. Observational
constraints on Barrow holographic dark energy were performed in
\cite{Ana}. A modified cosmological scenarios based on Barrow
entropy was presented in \cite{Emm2} which modifies the
cosmological field equations in such a way that contain new extra
terms acting as the role of an effective dark-energy sector. The
generalized second law of thermodynamics, when the entropy of the
Universe is in the form of Barrow entropy, was investigated in
\cite{Emm3}. Other cosmological consequences of the Barrow entropy
can be followed in \cite{Abr1,Mam,Abr2,Bar2,Sri,Das,Sha,Pra}.

The profound connection between gravitational field equations and
laws of thermodynamics has now been well established (see e.g.
\cite{Jac,Pad,Pad2,Pad3,Fro,Wang0,Ver,Cai5,CaiLM,sheyECFE,SheyLog,SheyPL,
Odin1} and references therein). It has been confirmed that gravity
has a thermodynamical predisposition and the Einstein field
equation of general relativity is just an equation of state for
the spacetime. Considering the spacetime as a thermodynamic
system, the laws of thermodynamics on the large scales, can be
translated as the laws of gravity. According to
``gravity-thermodynamics'' conjecture, one can rewrite the
Friedmann equations in the form of the first law of themodynamics
on the apparent horizon and vice versa \cite
{CaiKim,Cai2,Shey1,Shey2}. In line with studies to understand the
nature of gravity, Padmanabhan \cite{PadEm} argued that the
spacial expansion of our Universe can be understood as the
consequence of emergence of space. Equating the difference between
the number of degrees of freedom in the bulk and on the boundary
with the volume change, he extracted the Friedmann equation
describing the evolution of the Universe \cite{PadEm}. The idea of
emergence spacetime was also extended to Gauss-Bonnet, Lovelock
and braneworld scenarios \cite{CaiEm,Yang,Sheyem,Sheyem2,Sheyem3}.

In the present work, we are going to construct the cosmological
field equations of FRW universe with any special curvature, when
the entropy associated with the apparent horizon is in the form of
(\ref{S}). Our work differs from \cite{Emm2} in that the author of
\cite{Emm2} derived the modified Friedmann equations by applying
the first law of thermodynamics, $TdS =-dE$, to the apparent
horizon of a FRW universe with the assumption that the entropy is
given by (\ref{S}). Note that $-dE$ in \cite{Emm2} is just the
energy flux crossing the apparent horizon, and the apparent
horizon radius is kept fixed during an infinitesimal internal of
time $dt$. However, in the present work, we assume the first law
of thermodynamics on the apparent horizon in the form,
$dE=TdS+WdV$, where $dE$ is now the change in the energy inside
the apparent horizon due to the volume change $dV$ of the
expanding Universe. This is consistent with the fact that in
thermodynamics the work is done when the volume of the system is
changed. Besides, in \cite{Emm2}, the author focuses on a flat FRW
universe and modifies the total energy density in the Friedmann
equations by considering the contribution of the Barrow entropy in
the field equations, as a dark-energy component. Here, we consider
the FRW Universe with any special curvature and modify the
geometry (gravity) part (left hand side) of the cosmological field
equations based on Barrow entropy. The approach we present here is
more reasonable, since the entropy expression basically depends on
the geometry (gravity). For example, the entropy expressions
differ in Einstein, Gauss-Bonnet or $f(R)$ gravities. Any
modifications to the entropy should change the gravity (geometry)
sector of the field equations and vice versa. We shall also employ
the emergence idea of \cite{PadEm} to derive the modified
cosmological equations based on Barrow entropy. Again, we assume
the energy density (and hence the number of degrees of freedom in
the bulk) is not affected by the Barrow entropy, while the horizon
area (and hence the number of degrees of freedom on the boundary)
get modified due to the change in the entropy expression.
Throughout this paper we set $\kappa_B=1=c=\hbar$, for simplicity.

This paper is outlined as follows. In the next section, we derive
the modified Friedmann equations, based on Barrow entropy, by
applying the first law of thermodynamics of the apparent horizon.
In section \ref{GSL}, we examine the validity of the generalized
second law of thermodynamics for the universe filled with Barrow
entropy. In section \ref{Eme}, we derive the modified Friedmann
equations by applying the emergence scenario for the cosmic space.
We finish with conclusions in the last section.
\section{Modified Friedman Equations Based on Barrow entropy\label{FIRST}}
Our starting point is a spatially homogeneous and isotropic
universe with metric
\begin{equation}
ds^2={h}_{\mu \nu}dx^{\mu}
dx^{\nu}+\tilde{r}^2(d\theta^2+\sin^2\theta d\phi^2),
\end{equation}
where $\tilde{r}=a(t)r$, $x^0=t, x^1=r$, and $h_{\mu \nu}$=diag
$(-1, a^2/(1-kr^2))$ represents the two dimensional metric. The
open, flat, and closed universes corresponds to $k = 0, 1, -1$,
respectively. The boundary of the Universe is assumed to be the
apparent horizon with radius \cite{Sheyem}
\begin{equation}
\label{radius}
 \tilde{r}_A=\frac{1}{\sqrt{H^2+k/a^2}}.
\end{equation}
From the thermodynamical viewpoint  the apparent horizon is a
suitable horizon consistent with first and second law of
thermodynamics. Also, using the definition of the surface gravity,
$\kappa$, on the apparent horizon \cite{Cai2}, we can associate
with the apparent horizon a temperature which is defied as
\cite{Cai2,Sheyem}
\begin{equation}\label{T}
T_h=\frac{\kappa}{2\pi}=-\frac{1}{2 \pi \tilde
r_A}\left(1-\frac{\dot {\tilde r}_A}{2H\tilde r_A}\right).
\end{equation}
For $\dot {\tilde r}_A\leq 2H\tilde r_A$, the temperature becomes
$T\leq 0$. The negative temperature is not physically acceptable
and hence we define $T=|\kappa|/2\pi$. Also, within an
infinitesimal internal of time $dt$  one may assume $\dot {\tilde
r}_A\ll 2H\tilde r_A$, which physically means that the apparent
horizon radius is fixed. Thus there is no volume change in it and
one may define $T=1/(2\pi \tilde r_A )$ \cite{CaiKim}. The
profound connection between temperature on the apparent horizon
and the Hawking radiation has been disclosed in \cite{cao}, which
further confirms the existence of the temperature associated with
the apparent horizon.

The matter and energy content of the Universe is assumed to be in
the form of perfect fluid with energy-momentum tensor
\begin{equation}\label{T1}
T_{\mu\nu}=(\rho+p)u_{\mu}u_{\nu}+pg_{\mu\nu},
\end{equation}
where $\rho$ and $p$ are the energy density and pressure,
respectively. Independent of the dynamical field equations, we
propose the total energy content of the Universe satisfies the
conservation equation, namely, $\nabla_{\mu}T^{\mu\nu}=0$. This
implies that
\begin{equation}\label{Cont}
\dot{\rho}+3H(\rho+p)=0,
\end{equation}
where $H=\dot{a}/a$ is the Hubble parameter. Since our Universe is
expanding, thus we have volume change. The work density associated
with this volume change is defined as \cite{Hay2}
\begin{equation}\label{Work}
W=-\frac{1}{2} T^{\mu\nu}h_{\mu\nu}.
\end{equation}
For FRW background with stress-energy tensor (\ref{T1}), the work
density is calculated,
\begin{equation}\label{Work2}
W=\frac{1}{2}(\rho-p).
\end{equation}
We further assume the first law of thermodynamics on the apparent
horizon is satisfied and has the form
\begin{equation}\label{FL}
dE = T_h dS_h + WdV,
\end{equation}
where $E=\rho V$ is the total energy of the Universe enclosed by
the apparent horizon, and $T_{h}$ and $S_{h}$ are, respectively,
the temperature and entropy associated with the apparent horizon.
The last term in the first law is the work term due to change in
the apparent horizon radius. Comparing with the usual first law of
thermodynamics, $dE=TdS-pdV$, we see that the work term $-pdV$ is
replaced by $WdV$, unless for a pure de Sitter space where
$\rho=-p$, where the work term $WdV$ reduces to the standard
$-pdV$.

Taking differential form of the total matter and energy inside a
$3$-sphere of radius $\tilde{r}_{A}$, we find
\begin{equation}
\label{dE1}
 dE=4\pi\tilde
 {r}_{A}^{2}\rho d\tilde {r}_{A}+\frac{4\pi}{3}\tilde{r}_{A}^{3}\dot{\rho} dt.
\end{equation}
where we have assumed $V=\frac{4\pi}{3}\tilde{r}_{A}^{3}$ is the
volume enveloped by a 3-dimensional sphere with the area of
apparent horizon $A=4\pi\tilde{r}_{A}^{2}$. Combining with the
conservation equation (\ref{Cont}), we arrive at
\begin{equation}
\label{dE2}
 dE=4\pi\tilde
 {r}_{A}^{2}\rho d\tilde {r}_{A}-4\pi H \tilde{r}_{A}^{3}(\rho+p) dt.
\end{equation}
The key assumption here is to take the entropy associated with the
apparent horizon in the form of Barrow entropy (\ref{S}). The only
change needed is to replace the black hole horizon radius with the
apparent horizon radius, $r_{+}\rightarrow \tilde {r}_{A}$. If we
take the differential form of the Barrow entropy (\ref{S}), we get
\begin{eqnarray} \label{dS}
dS_h&=&
d\left(\frac{A}{A_0}\right)^{1+\delta/2}=\left(\frac{4\pi}{A_{0}}\right)^{1+\delta/2}
d\left(\tilde {r}_{A}^{2+\delta}\right)\nonumber\\
&=&(2+\delta)\left(\frac{4\pi}{A_{0}}\right)^{1+\delta/2}
 {\tilde
{r}_{A}}^{1+\delta} \dot{\tilde {r}}_{A} dt
\end{eqnarray}
Inserting relation (\ref{Work2}), (\ref{dE2}) and (\ref{dS}) in
the first law of thermodynamics (\ref{FL}) and using definition
(\ref{T}) for the temperature, after some calculations, we find
the differential form of the Friedmann equation as
\begin{equation} \label{Fried1}
\frac{2+\delta}{2\pi A_0 }\left(\frac{4\pi}{A_0}\right)^{\delta/2}
\frac{d\tilde {r}_{A}}{\tilde {r}_{A}^{3-\delta}}= H(\rho+p) dt.
\end{equation}
Combining with the continuity equation (\ref{Cont}), arrive at
\begin{equation} \label{Fried2}
-\frac{2+\delta}{2\pi A_0
}\left(\frac{4\pi}{A_0}\right)^{\delta/2} \frac{d\tilde
{r}_{A}}{\tilde {r}_{A}^{3-\delta}}=
 \frac{1}{3}d\rho.
\end{equation}
Integration yields
\begin{equation} \label{Fried3}
-\frac{2+\delta}{2\pi A_0
}\left(\frac{4\pi}{A_0}\right)^{\delta/2}\int{\frac{d\tilde
{r}_{A}}{\tilde {r}_{A}^{3-\delta}}} =  \frac{\rho}{3},
\end{equation}
which results
\begin{equation} \label{Frie3}
\frac{2+\delta}{2-\delta}\left(\frac{4\pi}{A_0}\right)^{\delta/2}
\frac{1}{2\pi A_0} \frac{1}{\tilde
{r}_{A}^{2-\delta}}=\frac{\rho}{3},
\end{equation}
where we have set the integration constant equal to zero. The
integration constant may be also regarded as the energy density of
the cosmological constant and hence it can be absorbed in the
total energy density $\rho$. Substituting $\tilde {r}_{A}$ from
Eq.(\ref{radius}) we immediately arrive at
\begin{equation} \label{Fried4}
\frac{2+\delta}{2-\delta}\left(\frac{4\pi}{{A_0}}\right)^{\delta/2}\frac{1}{2\pi
A_0
 } \left(H^2+\frac{k}{a^2}\right)^{1-\delta/2} = \frac{\rho}{3}
\end{equation}
 The above equation can be further rewritten as
\begin{equation} \label{Fried4}
\left(H^2+\frac{k}{a^2}\right)^{1-\delta/2} = \frac{8\pi G_{\rm
eff}}{3} \rho,
\end{equation}
where we have defined the effective Newtonian gravitational
constant as
\begin{equation}\label{Geff}
G_{\rm eff}\equiv \frac{A_0}{4} \left(
\frac{2-\delta}{2+\delta}\right)\left(\frac{A_0}{4\pi
}\right)^{\delta/2}.
\end{equation}
Equation (\ref{Fried4}) is the modified Friedmann equation based
on the Barrow entropy. Thus, starting from the first law of
thermodynamics at the apparent horizon of a FRW universe, and
assuming that the apparent horizon area has a fractal features,
due to the quantum-gravitational effects, we derive the
corresponding modified Friedmann equation of FRW universe with any
spatial curvature. It is important to note that in contrast to the
Friedmann equation derived in \cite{Emm2}, here the energy density
$\rho$ is not influenced by the Barrow entropy, while the effect
of the modified entropy contributes to the geometry sector of the
field equations. In the limiting case where $\delta=0$, the area
law of entropy is recovered and we have $A_{0}\rightarrow4G$. In
this case, $G_{\rm eff}\rightarrow G$, and Eq. (\ref{Fried4})
reduces to the standard Friedmann equation in Einstein gravity.

We can also derive the second Friedmann equation by combining the
first Friedmann equation (\ref{Fried4}) with the continuity
equation (\ref{Cont}). If we take the derivative of the first
Friedmann equation (\ref{Fried4}), we arrive at
\begin{equation}
2H\left(1-\frac{\delta}{2}\right) \left(\dot{H}-
\frac{k}{a^2}\right)\left(H^2+\frac{k}{a^2}\right)^{-\delta/2}=\frac{8
\pi G_{\rm eff}}{3} \dot{\rho}. \label{2Fri1}
\end{equation}
Using the continuity equation (\ref{Cont}), we arrive at
\begin{eqnarray}\label{2Fri2}
&&\left(1-\frac{\delta}{2}\right) \left(\dot{H}-
\frac{k}{a^2}\right)\left(H^2+\frac{k}{a^2}\right)^{-\delta/2}\nonumber\\
&&=-4\pi G_{\rm eff} (\rho+p).
\end{eqnarray}
Now using the  fact that $\dot{H}=\ddot{a}/a-H^2$, and replacing
$\rho$ from the first Friedmann equation (\ref{Fried4}), we can
rewrite the above equation as
\begin{eqnarray}
&&\left(1-\frac{\delta}{2}\right)
\left(\frac{\ddot{a}}{a}-H^2-\frac{k}{a^2}\right)\left(H^2+\frac{k}{a^2}\right)^{-\delta/2}\nonumber
\\
&&=-4\pi G_{\rm eff} p
-\frac{3}{2}\left(H^2+\frac{k}{a^2}\right)^{1-\delta/2}.\label{2Fri3}
\end{eqnarray}
After some simplification and rearranging terms, we find
\begin{eqnarray}
&&(2-\delta)\frac{\ddot{a}}{a}
\left(H^2+\frac{k}{a^2}\right)^{-\delta/2}+(1+\delta)\left(H^2+\frac{k}{a^2}\right)^{1-\delta/2}
\nonumber
\\
&&=-8\pi G_{\rm eff} p .\label{2Frie3}
\end{eqnarray}
This is the second modified Friedmann equation governing the
evolution of the Universe based on Barrow entropy. For $\delta=0$
($G_{\rm eff}\rightarrow G$), Eq. (\ref{2Frie3}) reproduces the
second Friedmann equation in standard cosmology
\begin{eqnarray}
2 \frac{\ddot{a}}{a}+H^2+\frac{k}{a^2}=-8\pi G p.\label{2Fri4}
\end{eqnarray}
If we combine the first and second modified Friedmann equations
(\ref{Fried4}) and (\ref{2Frie3}), we can obtain the equation for
the second time derivative of the scale factor. We find
\begin{eqnarray}
&&(2-\delta)\frac{\ddot{a}}{a}\left(H^2+\frac{k}{a^2}\right)^{-\delta/2}=-\frac{8\pi
G_{\rm eff}}{3} \left[(1+\delta) \rho +3p\right]\nonumber\\
&&=-\frac{8\pi G_{\rm eff}}{3}  \rho\left[(1+\delta) +3 w \right],
\label{2Fri5}
\end{eqnarray}
where $w=p/\rho$ is the equation of state parameter. Taking into
account the fact that $0\leq\delta\leq1$, the condition for the
cosmic accelerated expansion ($\ddot{a}>0$), implies
\begin{eqnarray}
 (1+\delta) +3 w<0  \  \     \longrightarrow   \  \  w< -\frac{(1+\delta)}{3}.          \label{w1}
\end{eqnarray}
When $\delta=0$, which corresponds to the simplest horizon
structure with area law of entropy, we arrive at the well-known
inequality $w<-1/3$ in Friedmann cosmology, while for $\delta=1$,
which implies the most intricate and fractal structure, we find
$w<-2/3$. This implies that, in an accelerating universe, the
fractal structure of the apparent horizon enforces the equation of
state parameter to become more negative.

In summary, in this section we derived the modified cosmological
equations given by Eqs. (\ref{Fried4}) and (\ref{2Fri5}) in Barrow
cosmology. These equations describe the evolution of the universe
with any spacial curvature, when the entropy associated with the
apparent horizon get modified due to the quantum-gravitational
effects. We leave the cosmological consequences of the obtained
Friedmann equations for future studies, and in the remanning part
of this paper, we focus on the generalized second law of
thermodynamics as well as derivation of Friedmann equation
(\ref{Fried4}) from emergence perspective.

\section{Generalized Second law of thermodynamics\label{GSL}}
Our aim here is to investigate another law of thermodynamics, when
the horizon area of the Universe has a fractal structure and the
associated entropy is given by Barrow entropy (\ref{S}). To do
this, we consider the generalized second law of thermodynamics for
the Universe enclosed by the apparent horizon. Our approach here
differs from the one presented in \cite{Emm3}. Indeed the authors
of \cite{Emm3} modified the total energy density in the Friedmann
equations based on Barrow entropy. The cosmological field
equations given in relations $(3)$ and $(4)$ of \cite{Emm3} are
nothing but the standard Friedmann equations, in a flat universe,
with additional energy component which acts as a dark energy
sector \cite{SheyCQ}. However, as we mentioned in the
introduction, here the effects of the Barrow entropy enter the
gravity (geometry) part of the cosmological field equations. Thus,
we assume the energy component of the Universe is not affected by
the Barrow entropy. Besides we consider the FRW universe with any
special curvature, while the authors of \cite{Emm3} only
considered a flat universe. In the context of the accelerating
Universe, the generalized second law of thermodynamics has been
explored in \cite{wang1,wang2,SheyGSL}.

Combining Eq. (\ref{Fried2}) with Eq. (\ref{Cont}) and using
(\ref{Geff}), we get
\begin{equation} \label{GSL1}
\frac{1}{\tilde {r}_{A}^{3-\delta}} (2-\delta) \dot{\tilde
{r}}_{A} =8\pi G_{\rm eff} H (\rho+p).
\end{equation}
Solving for $\dot{\tilde {r}}_{A}$, we find
\begin{equation} \label{dotr1}
\dot{\tilde {r}}_{A}=\frac{8\pi G_{\rm eff}}{2-\delta} H {\tilde
{r}_{A}^{3-\delta}}(\rho+p).
\end{equation}
Since $\delta\leq 1$, thus the sign of $\rho+p$  determines the
sign of $\dot{\tilde {r}}_{A}$. In case where the dominant energy
condition holds, $\rho+p\geq0$, we have
$\dot{\tilde{r}}_{A}\geq0$. Our next step is to calculate $T_{h}
\dot{S_{h}}$,
\begin{eqnarray}\label{TSh1}
T_{h} \dot{S_{h}}&=&\frac{1}{2\pi \tilde r_A}\left(1-\frac{\dot
{\tilde r}_A}{2H\tilde r_A}\right)\frac{d}{dt}
\left(\frac{A}{A_0}\right)^{1+\delta/2}\nonumber \\
&=& \frac{2+\delta}{2\pi}\left(1-\frac{\dot {\tilde r}_A}{2H\tilde
r_A}\right) \left(\frac{4\pi}{A_0}\right)^{1+\delta/2} {\tilde
{r}_A}^{\delta} \dot{\tilde {r}}_{A}
\end{eqnarray}
Substituting $\dot{\tilde {r}}_{A}$ from Eq. (\ref{dotr1}) and
using definition (\ref{Geff}), we reach
\begin{equation}\label{TSh2}
T_{h} \dot{S_{h}} =4\pi H {\tilde{r}_{A}^3}
(\rho+p)\left(1-\frac{\dot {\tilde r}_A}{2H\tilde r_A}\right).
\end{equation}
For an accelerating universe, the equation of state parameter can
cross the phantom line ($w=p/\rho<-1$), which means that the
dominant energy condition may violate, $\rho+p<0$. As a result,
the second law of thermodynamics, $\dot{S_{h}}\geq0$, no longer
valid. In this case, one can consider the total entropy of the
universe as, $S=S_h+S_m$, where $S_m$ is the entropy of the matter
field inside the apparent horizon. Therefore, one should study the
time evolution of the total entropy $S$. If the generalized second
law of thermodynamics holds, we should have
$\dot{S_{h}}+\dot{S_{m}}\geq0$, for the total entropy.

From the Gibbs equation we have \cite{Pavon2}
\begin{equation}\label{Gib2}
T_m dS_{m}=d(\rho V)+pdV=V d\rho+(\rho+p)dV,
\end{equation}
where here $T_{m}$ stands for the temperature of the matter fields
inside the apparent horizon. We further propose the thermal system
bounded by the apparent horizon remains in equilibrium  with the
matter inside the Universe. This is indeed the local equilibrium
hypothesis, which yields the temperature of the matter field
inside the universe must be uniform and the same as the
temperature of its boundary, $T_m=T_h$ \cite{Pavon2}. In the
absence of local equilibrium hypothesis, there will be spontaneous
heat flow between the horizon and the bulk fluid which is not
physically acceptable for our Universe. Thus, from the Gibbs
equation (\ref{Gib2}) we have
\begin{equation}\label{TSm2}
T_{h} \dot{S_{m}} =4\pi {\tilde{r}_{A}^2}\dot {\tilde
r}_A(\rho+p)-4\pi {\tilde{r}_{A}^3}H(\rho+p).
\end{equation}
Next, we examine the generalized second law of thermodynamics,
namely, we study the time evolution of the total entropy $S_h +
S_m$. Adding equations (\ref{TSh2}) and (\ref{TSm2}),  we get
\begin{equation}\label{GSL2}
T_{h}( \dot{S_{h}}+\dot{S_{m}})=2\pi{\tilde r_A}^{2}(\rho+p)\dot
{\tilde r}_A.
\end{equation}
Substituting $\dot {\tilde r}_A$ from Eq. (\ref{dotr1}) into
(\ref{GSL2}) we reach
\begin{equation}\label{GSL3}
T_{h}( \dot{S_{h}}+\dot{S_{m}})=\frac{16 \pi^2 }{2-\delta}G_{\rm
eff} H {\tilde r_A}^{5-\delta}(\rho+p)^2\geq0,
\end{equation}
which is clearly a non-negative function during the history of the
Universe. This confirms the validity of the generalized second law
of thermodynamics for a universe with fractal boundary, namely
when the associated entropy with the apparent horizon of the
universe is in the form of Barrow entropy (\ref{S}).
\section{Emergence of modified Friedmann equation \label{Eme}}
In his proposal \cite{PadEm}, Padmanabhan argued that gravity is
an emergence phenomena and the cosmic space is emergent as the
cosmic time progressed. He argued that the difference between the
number of degrees of freedom on the holographic surface and the
one in the emerged bulk, is proportional to the cosmic volume
change. In this regards, he extracted successfully the Friedmann
equation governing the evolution of the Universe with zero spacial
curvature \cite{PadEm}. In this perspective the spatial expansion
of our Universe can be regarded as the consequence of emergence of
space and the cosmic space is emergent, following the progressing
in the cosmic time. According to Padmanabhan's proposal, in an
infinitesimal interval $dt$ of cosmic time, the increase $dV$ of
the cosmic volume, is given by \cite{PadEm}
\begin{equation}
\frac{dV}{dt}=G(N_{\mathrm{sur}}-N_{\mathrm{bulk}}), \label{dV}
\end{equation}
where $G$ is the Newtonian gravitational constant. Here
$N_{\mathrm{sur}}$ and $N_{\mathrm{bulk}}$ stand for the number of
degrees of freedom on the boundary and in the bulk, respectively.
Following Padmanabhan, the studies were generalized to
Gauss-Bonnet and Lovelock gravity \cite{CaiEm}. While the authors
of \cite{CaiEm} were able to derive the Friedmann equations with
any spacial curvature in Einstein gravity, they failed to extract
the Firedmann equations of a nonflat FRW universe in higher order
gravity theories \cite{CaiEm}. In \cite{Sheyem}, we modified
Padmanabhan's proposal in such a way that it could produce the
Friedmann equations in higher order gravity theories, such as
Gauss-Bonnet and Lovelock gravities, with any spacial curvature.
The modified version of relation (\ref{dV}) is given by
\cite{Sheyem}
\begin{equation}
\frac{dV}{dt}=G\frac{\tilde{r}_A}{H^{-1}}
\left(N_{\mathrm{sur}}-N_{\mathrm{bulk}}\right). \label{dV1}
\end{equation}
Comparing with the original proposal of Padmanabhan in Eq.
(\ref{dV}), we see that in a nonflat universe, the volume increase
is still proportional to the difference between the number of
degrees of freedom on the apparent horizon and in the bulk, but
the function of proportionality is not just a constant, and is
equal to the ratio of the apparent horizon and Hubble radius. For
flat universe, $\tilde{r}_A =H^{-1}$, and one recovers the
proposal (\ref{dV}).

Our aim here is to derive the modified Friedmann equation based on
Barrow entropy from emergence proposal of cosmic space. Inspired
by Barrow entropy expression (\ref{S}), let us define the
effective area of the apparent horizon, which is our holographic
screen, as
\begin{eqnarray}
\widetilde{A} = A^{1+\delta/2}= \left(4 \pi
{\tilde{r}_A^2}\right)^{1+\delta/2}.
\end{eqnarray}
Next, we calculate the increasing in the effective volume as
\begin{eqnarray}\label{dVt1}
\frac{d\widetilde{V}}{dt}&=&\frac{\tilde{r}_A}{2}\frac{d\widetilde{A}}{dt}=\frac{2+\delta}{2}
(4 \pi \tilde{r}_A^{2})^{1+\delta/2} \dot{\tilde{r}}_A\nonumber\\
&=&\frac{1}{2}\ \frac{\delta+2}{\delta-2}(4 \pi)^{1+\delta/2}
\tilde{r}_A^5 \ \frac{d}{dt}\left(\tilde{r}_A ^{\delta-2}\right).
\label{dVt2}
\end{eqnarray}
Our first key assumption here is to specify the correct expression
for the number of degrees of freedom on the apparent horizon,
$N_{\mathrm{sur}}$. Motivated by (\ref{dVt2}) and following
\cite{Sheyem}, we choose
\begin{eqnarray}
N_{\mathrm{sur}}&=&\frac{4\pi {\tilde{r}_A}^{2+\delta}}{G_{\rm
eff}} , \label{Nsur2}
\end{eqnarray}
where we have used (\ref{Geff}). We also assume the temperature
associated with the apparent horizon is the Hawking temperature,
which is given by \cite{CaiEm}
\begin{equation}\label{T2}
T=\frac{1}{2\pi \tilde{r}_A},
 \end{equation}
and the energy contained inside the sphere with volume $V=4
\pi\tilde{r}^3_A/3$ is the Komar energy
\begin{equation}
E_{\mathrm{Komar}}=|(\rho +3p)|V.  \label{Komar}
\end{equation}
Employing the equipartition law of energy, we can define the bulk
degrees of freedom as
\begin{equation}
N_{\mathrm{bulk}}=\frac{2|E_{\mathrm{Komar}}|}{T}.  \label{Nbulk}
\end{equation}
In order to have $N_{\rm bulk}>0$, we take $\rho+3p<0$
\cite{PadEm}. Thus the number of degrees of freedom in the bulk is
obtained as
\begin{equation}
N_{\rm bulk}=-\frac{16 \pi^2}{3}  \tilde{r}_A^{4} (\rho+3p),
\label{Nbulk}
\end{equation}
The second key assumption here is to take the correct form of
expression (\ref{dV1}). To write the correct proposal, we make
replacement $G\rightarrow \Gamma^{-1}$ and $V\rightarrow
\widetilde{V}$ in the proposal (\ref{dV1}) and rewrite it as
\begin{equation}\label{dV2}
\Gamma \frac{d\widetilde{V}}{dt}=\frac{\tilde{r}_A}{H^{-1}}
(N_{\mathrm{sur}}-N_{\mathrm{bulk}}).
\end{equation}
where $\Gamma={4}/{A_0^{1+\delta/2}}$. Substituting relations
(\ref{dVt1}), (\ref{Nsur2}) and (\ref{Nbulk}) in Eq. (\ref{dV2}),
after simplifying, we arrive at
\begin{eqnarray}
&&\frac{4}{A_0} \frac{2+\delta}{2}\left(\frac{4
\pi}{A_0}\right)^{\delta/2}\tilde{r}_A ^{2+\delta}\dot{\tilde{r}}_{A}\nonumber\\
&=&\frac{\tilde{r}_A}{H^{-1}}
\left[\frac{{\tilde{r}_A}^{2+\delta}}{G_{\rm eff}} +\frac{4
\pi}{3}  \tilde{r}_A^{4} (\rho+3p)\right].  \label{Frgb1}
\end{eqnarray}
Using definition (\ref{Geff}), the above equation can by further
simplified as
\begin{eqnarray}
(2-\delta){\tilde{r}_A}^{\delta-3}\frac{\dot{\tilde{r}}_A}{H}-2\
{\tilde{r}_A}^{\delta-2}=\frac{8\pi G_{\rm eff}}{3}(\rho+3p).
\label{Frgb11}
\end{eqnarray}
If we multiply the both side of Eq. (\ref{Frgb11}) by factor
$2\dot{a}a$, after using the continuity equation (\ref{Cont}), we
arrive at
\begin{equation}
\frac{d}{dt}\left( a^2 \tilde{r}_A^{\delta-2}\right)=\frac{8 \pi
G_{\rm eff}}{3} \frac{d}{dt}(\rho a^2). \label{Frgb2}
\end{equation}
Integrating, yields
\begin{equation}
\left(H^2+\frac{k}{a^2}\right)^{1-\delta/2}=\frac{8 \pi G_{\rm
eff}}{3} \rho, \label{Frgb3}
\end{equation}
where in the last step we have used relation (\ref{radius}), and
set the integration constant equal to zero. This is indeed the
modified Friedmann equation derived from emergence of cosmic space
when the entropy associated with the apparent horizon is in the
form of Barrow entropy (\ref{S}). The result obtained here from
the emergence approach coincides with the obtained modified
Friedmann equation from the first law of thermodynamics in section
\ref{FIRST}. Our study indicates that the approach presented here
is enough powerful and further supports the viability of the
Padmanabhan's perspective of emergence gravity and its
modification given by Eq. (\ref{dV2}).
\section{conclusions \label{Con}}
Recently, and motivated by the Covid-19 virus structure, Barrow
proposed a new expression for the black hole entropy
\cite{Barrow}. He demonstrated that taking into account the
quantum-gravitational effects, may lead to intricate, fractal
features of the black hole horizon. This complex structure implies
a finite volume for the black hole but with infinite/finite area
for the horizon. In this viewpoint, the deformed entropy
associated with the black hole horizon no longer obeys the area
law and increases compared to the area law due to fractal
structure of the horizon. The amount of increase in entropy
depends on the amount of quantum-gravitational deformation of the
horizon which is characterized by an exponent $\delta$.

Based on Barrow's proposal for black hole entropy and assuming the
entropy associated with the apparent horizon of the Universe has
the same expression as black hole entropy, we investigated the
corrections to the Firedmann equations of FRW universe, with any
spacial curvature. These corrections come due to the
quantum-gravitational fractal intricate structure of the apparent
horizon. To do this, and motivated by the
``gravity-thermodynamics'' conjecture, we proposed the first law
of thermodynamics, $dE=T_hdS_h+WdV$, holds on the apparent horizon
of FRW universe and the entropy associated with the apparent
horizon is given by Barrow entropy (\ref{S}). Starting from the
first law of thermodynamics and taking the entropy in the form of
Barrow entropy (\ref{S}), we extracted modified Friedmann
equations describing the evolution of the Universe. Then, we
checked the validity of the generalized second law of
thermodynamics by considering the time evolution of the matter
entropy together with the Barrow entropy associated with the
apparent horizon. We also employed the idea of emergence gravity
suggested by Padmanabhan \cite{PadEm} and calculated the number of
degrees of freedom in the bulk and on the boundary of universe.
Subtracting $N_{\rm sur}$ and $N_{\rm bulk}$ and using the
modification of Padmanabhan's proposal given in Eq. (\ref{dV2}),
we were able to extract Friedmann equations which is modified due
to the presence of Barrow entropy. This result coincides with the
one obtained from the first law of thermodynamics. Our study
confirms the viability of the emergence gravity proposed in
\cite{PadEm,Sheyem}.

Many interesting topics remain for future considerations. The
cosmological implications of the modified Friedmann equations and
the evolution of the Universe can be addressed. The influences of
the modified Friedmann equations on the gravitational collapse,
structure formation and galaxies evolution can be investigated.
The effects of the fractal parameter $\delta$ on the thermal
history of the Universe, as well as anisotropy of CMB are also of
great interest which deserve exploration.  These studies belong
beyond the scope of the present work and we leave them for the
future projects.

\acknowledgments{I thank Shiraz University Research Council.}


\end{document}